\documentclass[twocolumn,preprintnumbers,showpacs,amsmath,amssymb,amsthm,aps,prd,nofootinbib]{revtex4-1}
\usepackage{graphicx}
\usepackage{color}
\usepackage[utf8]{inputenc}
\usepackage[colorlinks,hyperindex]{hyperref}  
\usepackage{accents}
\usepackage{enumitem}
\definecolor{byzantine}{rgb}{0.74, 0.2, 0.64}
\definecolor{vividviolet}{rgb}{0.62, 0.0, 1.0}
\hypersetup
 {
   colorlinks,%
   citecolor=blue,%
   linkcolor=blue,%
   urlcolor=blue,%
 }
 
\def\setR{\mathbb{R}}

\newcommand{\sss}[1]{\scriptscriptstyle #1} 

\newcommand{\pbundle}[4]{#1(#2,#3,#4)} 
\newcommand{\sbundle}[3]{#1(#2,#3)} 

\newtheorem{thm}{Theorem}[section]

\begin{document}

\title[TEGR as gauge theory: coupling with Cartan connection]{Teleparallel gravity  as a gauge theory: coupling to matter with Cartan connection}

\author{ E.~Huguet$^1$, M.~Le~Delliou$^{2,3}$, M.~Fontanini$^1$, and Z.-C. Lin$^4$}
\affiliation{$1$ - Universit\'e de Paris, APC-Astroparticule et Cosmologie (UMR-CNRS 7164), 
 F-75006 Paris, France.}
\email{michele.fontanini@gmail.com\\
huguet@apc.univ-paris7.fr}
\affiliation{$2$ - Institute of Theoretical Physics, School of Physical Science and Technology, Lanzhou University,
No.222, South Tianshui Road, Lanzhou, Gansu 730000, P R China
}
\affiliation{$3$ - Instituto de Astrof\'isica e Ci\^encias do Espa\c co, Universidade de Lisboa,
Faculdade de Ci\^encias, Ed.~C8, Campo Grande, 1769-016 Lisboa, Portugal}
\email{(delliou@lzu.edu.cn,)morgan.ledelliou.ift@gmail.com}
\affiliation{$4$ - Institute of Theoretical Physics \& Research Center of Gravitation, Lanzhou University, China
}
\email{linzch12@lzu.edu.cn}

\date{\today}
\begin{abstract}
We present a consistent and complete description of the coupling to matter in the Teleparallel Equivalent to General Relativity (TEGR) theory built from a Cartan connection, as we proposed in previous works. 
A first theorem allows us to obtain parallel transport from the Cartan connection into a proper Ehresmann connection, while a second ensures to link the TEGR-Cartan connection to the Ehresmann one-form that contains the Levi-Civita connection. This yields a coupling to matter in agreement with observations and the Equivalence Principle. As the fundamental fields proceed from the Cartan connection, if one insists on interpreting TEGR as a gauge theory of translations, such translation gauge field can be extracted from the consistent theory presented. However, this would entail a fundamental change in the structures known for gauge theory and a split between gauge field and connection is imperative. The willingness to take such a step is left to the reader.
\end{abstract}
\maketitle
\tableofcontents
\section{Introduction}\label{SEC-Intro}

In this paper we describe how a coupling to matter can be obtained in a new mathematical framework for the  Teleparallel Equivalent to General Relativity (TEGR) using a reductive Cartan connection.
We start from the Levi-Civita connection, or its corresponding one-form in the Cartan (tetrads) formalism, to describe the coupling to matter fields, which, from observational grounds, is known to describe it well.

TEGR is the well known theory in which the effect of gravity does not manifest itself, as in General Relativity (GR), through  induced 
curvature to spacetime, but by instead  giving it torsion. Although being equivalent to GR, this approach yields both an alternative description and 
interpretation to gravity. It has been presented with various perspectives: in Ref.~\cite{Aldrovandi:2013wha} for its translation-gauge presentation, in Ref.~\cite{Maluf:2011kf} for the ``pure tetrads formalism" approach, while in ~Ref.~\cite{BeltranJimenez:2019tjy}, tensorial formalism is used to present it, together with another gravitation equivalent 
to GR, first introduced in Ref.~\cite{Nester:1998mp}, the Symmetric TEGR (STGR). TEGR has also provided a stout base for modified gravities, such as $f(T)$ \cite{Ferraro:2006jd,Capozziello:2019cav} theories, further generalised with $f(R, T)$ \cite{Bahamonde:2015zma}, $f(T,B)$ \cite{Bahamonde:2016grb}. Additional important generalisation comprise Conformal TEGR \cite{Maluf:2011kf, Formiga:2019frd,Bamba:2013jqa}, or Teleparallel Equivalent to Lovelock Gravity \cite{Gonzalez:2019tky}, which are also actively studied. 

Although the familiar tensorial formalism can be used to formulate TEGR \cite[as in][]{BeltranJimenez:2019tjy}, its presentation as a gauge theory for the translation group \cite[see][and references herein]{Aldrovandi:2013wha,Krssak:2018ywd} allows one to obtain the torsion 
as the curvature of a connection 
defined in the principal bundle of translations. Recently, some of the authors pointed out a difficulty
in this formalism: the connection one-form is implicitly identified with another mathematical object required to define the torsion, the 
so-called canonical one-form. This one-form is only defined in the bundle of frames and is not a connection, two properties
turning the identification problematic \cite{Fontanini:2018krt,Pereira:2019woq,LeDelliou:2019esi}.

The idea of relating TEGR, and, more generally, gravity, to translations is a very natural one, since the Noether current associated 
with them is the energy-momentum tensor. On another hand, extending that link to a gauge theory of the translations group could be
more questionable. Indeed, gauge theories are very successful in the standard model of particle physics describing fundamental 
interactions, except gravity, in Minkowski space. There, the gauges groups ($U(1)$, $SU(2)$, {\it etc...}) only act on fields, not on 
spacetime (their action is often said to be ``internal"). 
The aim of these theories is to implement the local invariance of a matter equation (usually the Dirac equation), under their respective 
symmetry group. This is achieved by introducing gauge fields which couple minimally to matter fields. In that context, local 
invariance means that the action of the symmetry group on matter fields depends on the position in spacetime. 

In gravity the group of invariance involved is the group of local diffeomorphisms  of $\setR^4$ (GL$(4,\setR)$), that is the changes of coordinates mapping spacetime. 
Its appearence is motivated by the Equivalence Principle (EP) translated operationally into the General Covariance Principle (GCP). In the Cartan 
(tetrads) formulation, any spacetime tensor is mapped by tetrads to a Lorentz tensor.\footnote{This corresponds to the usual change of indices: 
$T^{\ldots a \ldots}_{\ldots b \ldots}(x) = e^a_\mu(x) e^\nu_b(x) T^{\ldots \mu \ldots}_{\ldots \nu \ldots}(x)$.} Thus, all fields, except the tetrads 
themselves, are viewed as scalars under changes of coordinates. These fields also belong to representations of the Lorentz group. Indeed, using the appropriate representation of the Lorentz group, spinorial fields can also be taken into 
account, which is a central feature of the tetrad approach. In this formalism the GCP translates into two invariances: the invariance as a scalar under the change of coordinates, and the invariance under 
the local Lorentz transformations in the corresponding representation \cite{Weinberg:1972}.
In this tetrad framework, the coupling of matter fields to gravity is usually obtained by the minimal coupling prescription, where the partial derivative are replaced by the covariant Fock-Ivanenko derivative.
In this derivative the coupling term is the spin (or Lorentz) connection, a one-form valued in the Lorentz Lie algebra $\mathfrak{so}(1,3)$. Then, 
from a gauge theory perspective, if that coupling can be associated with a local Lorentz invariance, the relation with the local diffeomorphism 
invariance (already satisfied) is not that obvious. Moreover, the status of tetrads, as representing the gravitational field, is, in a gauge theory framework, unclear. 

Indeed, a large amount of works since the sixties has been done to clarify this situation\footnote{To illustrate how that point has been already recognized since the birth of gauge theories let us quote the introduction of 
a review of 1985 by Ivanenko and Sardanashvily \cite[p. 4]{Ivanenko:1984vf}: "The main dilemma which during 25 years has been confronting the establishment of the gauge
gravitation theory, is that gauge potentials represents connections on fiber bundles, while gravitational fields in GR
are only metric or tetrad (vierbein) fields." }.  Different theories, using different symmetry group 
($\setR^4$, Poincar\'e, GL$(4,\setR)\rtimes \setR^4$, SO$(1,4)$, \ldots), not limited to the diffeomorphism group (GL$(4,\setR)$), have been obtained. A comprehensive review of this gauge approach is \cite{Blagojevic:2013xpa}.

We will not directly address here this gauge problem in its generality. 
Instead, we will describe a mathematical framework for TEGR in which the diffeomorphism invariance is canonically 
satisfied. In this framework, the minimal coupling, through the Levi-Civita connection one-form, is consistent with current observations, 
and derives from 
a specific Cartan connection, chosen such that curvature is the torsion. In the tetrad formalism, this amounts to retrieving the usual Fock-Ivanenko covariant derivative. Such derivation from a connection is  suitable from a gauge theory perspective of TEGR. This  basis will enable us to map out the modifications required on the established framework of gauge theories, if we insist on interpreting our result as a gauge theory for translations: mainly, the dissociation of the gauge field from a connection, and the restriction to the translation ``gauge'' group, appearing only through its algebra.

The structure of the paper is as follows:  Sec.~\ref{SEC-WhichConnec} reviews the motivations in the choice of a reductive Cartan connection; 
 Sec.~\ref{SEC-CouplingOverview} summarizes the relevant issues involved in coupling gravity to matter with a Cartan connection and gives a setup 
for the derivation of that coupling to matter; technical details are the subject of Sec.~\ref{SEC-CoupMatDetails}; we then discuss 
Sec.~\ref{SEC-TEGRAsTrans}, the extensions of the gauge paradigm which could be considered in order to interpret TEGR, obtained from our 
results, as a gauge theory for translations; we finally conclude Sec.~\ref{SEC-Conclu}; some complements on technical details are given in 
appendices.

For general notions and definitions regarding differential geometry
we refer the Reader to \cite{Fecko:2006, Isham:1999qu, Nakahara:2003, KobayashiNomizu:1963}


\section{Which connection for TEGR?}\label{SEC-WhichConnec}

In this section we remind our motivation in using a Cartan connection to describe TEGR. For the sake of completeness let us first recall some defining properties of the Cartan connection. More detailed account in the context of gravity 
may be found in \cite{Wise_2010, Catren:2014vza}, while a comprehensive mathematical reference is \cite{Sharpe:1997}.

\subsection{About Cartan connection}\label{SUBSEC-CartanConnec}
Let us first recall some facts about Ehresmann connections on principal $G$-bundle (a principal fiber bundle of Lie group $G$). 
Each tangent space on a point of the fiber bundle contains a vertical subspace defined as the tangent space of the fiber at this point. Any  complementary
subspace of this vertical space is said to be horizontal. An Ehresmann connection defines in a unique way the notion of horizontality in a principal $G$-bundle: it specifies horizontal subspaces. This is usually done through a connection one-form $\omega_{\sss E}$ whose kernel defines a horizontal space at each point of the total space. Such a one-form is defined by the following properties:
\begin{enumerate}
\item it takes its values in the Lie algebra $\mathfrak{g}$ of the Lie group $G$,
\item it is G-invariant: $R_g^* \omega = Ad_{g^{-1}}\omega$, $R_g$ being the right action of $G$ on the bundle,
\item it reduces to the Maurer-Cartan form $\omega_{\sss G}$ of the group $G$ along the fibers: 
$\omega_{\sss E}(V) = \omega_{\sss G} (V)$, for any vertical vector $V$. 
\end{enumerate}

Now, let us recall the definition properties of a Cartan connection on a principal $H$-bundle, $H$ being a subgroup of a Lie group $G$. 
The Cartan connection is defined through the one form $\omega_{\sss C}$ such that:
\begin{enumerate}
\item it takes values in the algebra $\mathfrak{g}\supset\mathfrak{h}$ of $G \supset H$.\label{enu:Gval}
\item it is H-invariant: $R_h^* \omega = Ad_{h^{-1}}\omega$, $R_h$ being the right action of $H$ on the bundle,
\item it reduces to the Maurer-Cartan form $\omega_{\sss H}$ of the group $H$ along the fibers,
\item it is, at each point $p$ of the $H$-bundle, a linear isomorphism between the tangent space $T_pP$ at $p$ and the Lie algebra $\mathfrak{g}$. 
This property requires that $G$ has the same dimension as the tangent space $T_p P$.\label{enu:gConn}
\end{enumerate}
The properties (\ref{enu:Gval}) and (\ref{enu:gConn}) distinguish the Cartan connection from Ehresmann's. As a consequence
of the above properties, the tangent space of the base manifold $M$ can be locally 
identified with the tangent space $\mathfrak{g}/\mathfrak{h}$ of the homogeneous 
space\footnote{
Note that both $G/H$, with $H$ a closed subgroup of $G$, being   
a homogeneous space, and the fact that $\mathfrak{g}/\mathfrak{h}$ 
can be identified with its tangent space are known results of differential geometry of Lie groups (see for instance \cite{Fecko:2006} p. 294  for the former statement, and \cite{Sharpe:1997} p. 163, for the latter).
}
$G/H$. Indeed, the condition (\ref{enu:gConn}) precisely states
that the $H$-principal bundle is soldered to its base $M$.

For a $(3+1)$-dimensional manifold there are only three possible homogeneous spaces: both (Anti)-de~Sitter spaces, and the Minkowski space.  The corresponding Cartan geometries have the property of being reductive  \cite[see][p197 for a definition]{Sharpe:1997}, implying
that, the Cartan connection takes the form
\begin{equation*}
 \omega_{\sss C} = \omega + \theta,
\end{equation*}
where $\omega$ is an $\mathfrak{h}$-valued connection one-form of  Ehresmann type,  and 
$\theta$ a $\mathfrak{g}/\mathfrak{h}$-valued one-form, both
defined on the principal $H$-bundle. The reductive property also imply that the curvature of the Cartan connection
splits into two parts: 
\begin{equation*}
\Omega(\omega_{\sss C}) = \Omega(\omega) + \Theta(\omega),
\end{equation*}
where $\Omega(\omega) = d\omega + \omega \wedge \omega$ is the curvature of the connection $\omega$ and $\Theta(\omega) = d\theta + \omega\wedge \theta$
its torsion.
\subsection{Cartan connection in TEGR}\label{SUBSEC-CartanConnecTEGR}

In previous works \cite{Fontanini:2018krt,LeDelliou:2019esi} we found that one encounters difficulties in the formulation of TEGR as a gauge theory 
of translations, mainly because of the problematic implicit identification of the Ehresmann translation connection, defined in a principal 
translation bundle ($\setR^4$-bundle),  with the canonical one-form $\theta$ appearing in the definition of torsion: $\Theta(\omega) = d\theta + \omega\wedge \theta$, and which is not an Ehresmann connection. The $\setR^4$-valued one-form  $\theta$  is defined in the bundle of frames $LM$,
a principal GL$(4,\setR)$-bundle, and its sub-bundle of orthonormal frames $OM$, a principal SO$(1,3)$-bundle, through
\begin{equation}\label{EQ-DefCanno}
 (\theta(e), V) = (e^{-1}, \pi_* V),
\end{equation}
where $e$ is a frame in $LM$  over a point $x$ of the base manifold $M$, $V$ a vector of $TLM$, and $\pi$ the projection on the base. 
In coordinates the above relation reads: $\theta^a(e)[V] = e^a[\pi_*V] = V^a$. 

In an attempt to remedy this situation, we proposed to consider 
a Cartan connection directly defined over the bundle of orthonormal frames $OM$.  The choice of a Cartan connection is first of all motivated 
by the fact that, when defined on $OM$, it can be chosen such that its curvature is the torsion, a central property for TEGR. This is related to the 
property for the Cartan connection on $OM$ to be reductive (see Sec. \ref{SUBSEC-CartanConnec}). In this case the curvature-less and torsion-full 
Cartan connection one-form reads 
\begin{equation}\label{EQ-CartanReducConnec-W+thet}
 \omega_{\sss C} = \omega_{\sss W} + \theta,   
\end{equation} 
where the Ehressman term, $\omega_{\sss W}$, is the curvature-less Weitzenb\"ock  one-form connection  
and where the term, $\theta$, coincides with the canonical one-form on $OM$.

We remark that the term $\theta$ of the Cartan connection (\ref{EQ-CartanReducConnec-W+thet}) takes its values 
precisely in the translation sub-algebra $\mathfrak{g}/\mathfrak{h} = \setR^4$ of the Poincar\'e algebra, thus local translations
are implemented in the Cartan connection although $\theta$ is not a connection by itself.


\section{The coupling to matter: overview}\label{SEC-CouplingOverview}
This section aims to summarize how we obtain the Levi-Civita covariant derivative from the reductive Cartan connection  
we proposed as a possible connection to describe  TEGR as a gauge theory.

The connections usually used in gauge theories are of Ehresmann type. They are associated to parallel transport and covariant
derivative. These two notions are not directly available for a Cartan connection. Therefore, to obtain a covariant derivative from 
a Cartan connection leads to associate it to an Ehresmann connection. An obvious way to realize this mapping in the case of a reductive 
Cartan connection, is to define  parallel transport through the Ehresmann part of the connection. In our context, this part 
is the Weitzenb\"ock connection $\omega_{\sss W}$, which obviously cannot give the Levi-Civita coupling\footnote{Which would respect observational evidence and the Equivalence Principle.}. We thus have to consider 
a different way to implement parallel transport. Remark that, the use of the Weitzenb\"ock connection for the parallel transport would imply
that the spacetime manifold be parallelizable -- which 
correspond to a trivial bundle of frames --  and thus the theory not strictly equivalent to GR\footnote{Note, however that the solutions 
exculded from GR by the parallelizability are discarded on physical basis \cite{Geroch:1968zm,Geroch:1970uv}.}. 
Another, more general, way to build a covariant derivative from a Cartan connection is to use a theorem (proven by R.~Sharpe in 
\cite{Sharpe:1997} and reproduced in Appendix \ref{App-EhressToCart}) which, essentially, gives a map between the set of Cartan connections 
and that of Ehresmann connections satisfying a technical condition (detailed in Sec.~\ref{SUBSEC-SharpeTheorem}). We will, in Sec.~\ref{SUBSEC-jModifContorsion}, make use of this theorem to map the Cartan connection $\omega_{\sss C}$ defined on the orthonormal frame bundle $OM$ to 
an affine connection on the principal Poincaré bundle $AM$ (the affine extention of $OM$ viewed as a principal Lorentz bundle). 
The specialization to the orthonormal bundle, which can be related to the Equivalence Principle,  allows us  to introduce the contorsion one-form in order 
to obtain a Levi-Civita connection as the Lorentz part of the affine (Poincaré) connection (this is detailed in 
Sec.~\ref{SUBSEC-jModifContorsion}). This affine connection is then mapped back onto the orthonormal frame bundle $OM$ where it divides
into two parts: the Levi-Civita connection $\omega_{\sss LC}$  and the canonical one-form $\theta$. However, 
the Levi-Civita connection obtained that way, appears as the combination of the Weitzenb\"ock connection and its related 
contorsion $\kappa_{\sss W}$, namely 
$\omega_{\sss LC} = \omega_{\sss W} - \kappa_{\sss W}$, a well known result, reformulated in the case of fiber bundle in 
Ref.~\cite[theorem 6.2.5 p. 79]{Bleecker:1981}. Thus, we finally recover the Levi-Civita coupling to matter under the 
form obtained in \cite{Aldrovandi:2013wha}, opening the possibility to relate this connection to (local) translations.  

\section{The coupling to matter: details}\label{SEC-CoupMatDetails}
In this section we detail our proposal to obtain the Levi-Civita covariant derivative from the reductive Cartan connection 
(\ref{EQ-CartanReducConnec-W+thet}). Our approach mainly relies on a theorem, hereafter referred as Sharpe theorem 
 relating Cartan and Ehresmann connections. This theorem is proved in \cite{Sharpe:1997} and reproduced in Appendix~\ref{App-EhressToCart}. We first examine this theorem, and then show how  to switch from the Weitzenb\"ock to the Levi-Civita
connection in the Lorentz sector. Finally, we use these properties to obtain the covariant derivative on the spacetime manifold.

\subsection{The Sharpe theorem}\label{SUBSEC-SharpeTheorem}
Let us first consider the main ingredients of the Sharpe theorem (theorem~\ref{THM-Sharpe}) and how it specializes in our context. The theorem 
is concerned by a principal $H$-bundle $\sbundle{P}{M}{H}$ and its so-called $G$-extension\footnote{see for instance \cite{Isham:1999qu} and 
Appendix \ref{App-AssocBundleQ}.}: the principal $G$-bundle $\sbundle{Q}{M}{G} = P \times_{\sss H} G$. 
In our framework, a bundle formalism for TEGR with a Cartan connection \cite{Fontanini:2018krt}, $P$ is the bundle of orthonormal frames $OM$ identified with the principal Lorentz($SO_0(1,3)$)-bundle, and $Q$ is the Poincar\'e-bundle $AM$, its corresponding affine bundle with structure group $\mathcal{P}\nolinebreak:=\nolinebreak SO_0(1,3) \rtimes \setR^4$. 

A central assumption of the theorem is the existence of an $H$ bundle-map. 
$\varphi: P \longrightarrow Q$, that is a continuous map such that $\varphi(p h) = \varphi(p) h$ with $p \in P$ and $h\in H \subset G$. 
The theorem states that $\varphi^*$ is a one-to-one correspondence between the set of Ehresmann connections on $Q$ (here $AM$),
whose kernel does not contains non-null vectors of $\varphi_*(TP)$ (the technical condition we mention in Sec.~\ref{SEC-CouplingOverview}), 
and the set of Cartan connections on $P$ (here $OM$). 

Our goal is to obtain explicitly the expression, on a Cartan connection one-form  $\omega_{\sss C}$, of the 
inverse map of $\varphi^*$, that is $\left(\varphi^*\right)^{-1}(\omega_{\sss C})$.
In theorem \ref{THM-Sharpe},  $\left(\varphi^*\right)^{-1}$ isn't formulated in a closed form. Instead, it is obtained as the extension of a general Cartan connection 
$\omega_{\sss C}$ to a one-form on the product $P \times G$: $j(\omega_{\sss C})$, in Eq.~(\ref{EQ-def-j}). This one-form $j(\omega_{\sss C})$ is 
proven to be the pull-up from $Q$ to $P \times G$ of an Ehresmann connection $\omega_{\sss E}$ whose kernel intersection with $\varphi_*(TP)$ 
is reduced to zero, that is precisely the image of $\omega_{\sss C}$ by $\left(\varphi^*\right)^{-1}$. 

An explicit expression of 
$\omega_{\sss E}\nolinebreak=\nolinebreak\left(\varphi^*\right)^{-1}(\omega_{\sss C})$ for a Cartan connection $\omega_{\sss C}$ requires to specify Eq.~(\ref{EQ-def-j}) in a local trivialization. 

We first recast it for our  matrix  Lie group:
\begin{equation}\label{EQ-j-matrix}
    j(\omega_{\sss C}) = g^{-1} \omega_{\sss C} g + g^{-1}dg,
\end{equation}
taking into account the projections appearing in Eq.~(\ref{EQ-def-j})  by recalling, for a product of manifolds $M \times N$, 
that one can always write $T_{(m,n)} (M \times N) = T_m M \oplus T_n N$. In this form, $\omega_{\sss C}$ acts on $TP$ and $g^{-1}dg$, the Maurer-Cartan form of $G$, acts on $TG$.

To simplify matters, let us specify a local trivialization  -- $f_a: \pi^{-1}(U_a) \longrightarrow U_a \times F$, where 
$\pi$ is the projection on the base, $\{U_a\}$ an open covering of $M$ and $F$ a fiber -- by the local product $U \times F$ for 
some open set $U$ of the covering. We then set 
\begin{align*}
    &P_{\sss U} := U \times H,\\
    &Q_{\sss U} := U \times G,
\end{align*}
and accordingly: $(P\times G)_{\sss U} = P_{\sss U} \times G$. Local coordinates for points $p \in P_{\sss U}$, $q \in Q_{\sss U}$ 
will be hereafter denoted by $p = (x,h)$ and $q=(x,g)$ respectively, with $x \in U$.

In a local trivialisation corresponding to $U\times G$, $G$ being a matrix Lie group, 
a connection one-form $\omega$ can be decomposed as:
\begin{equation}\label{EQ-OmegaLocalTriv}
    \omega(x,g) = g^{-1} \omega^{\sss U}(x) g + g^{-1}dg,
\end{equation}
where $\omega^{\sss U}(x)$ acts on tangent vectors of $T_x U \subset T_x M$ and $g^{-1}dg$ is the Maurer-Cartan form acting on vectors of
$T_g G$.  Applying Eq.~(\ref{EQ-OmegaLocalTriv}) to the Cartan connection $\omega_{\sss C}$ in the trivialisation corresponding to $P_{\sss U}$ gives
\begin{equation}\label{EQ-OmegaCartanTriv}
    \omega_{\sss C}(x,h) = h^{-1}\omega^{\sss U}_{\sss C}(x) h + h^{-1} dh.
\end{equation}

Inserting Eq.~(\ref{EQ-OmegaCartanTriv}) in Eq.~(\ref{EQ-j-matrix}) gives the expression of $j(\omega_{\sss C})$ in the local trivialization 
corresponding to $P_{\sss U} \times G$: 
\begin{equation}\label{EQ-j-matrix-Triv}
\begin{split}
   \left(j(\omega_{\sss C})\right) (x, h, g) & = g^{-1}
   \left( h^{-1}\omega^{\sss U}_{\sss C}(x) h + h^{-1} dh\right) g \\
   &+ g^{-1}dg,
\end{split}
\end{equation}

In the trivializations defined above, the mapping from $P \times G$ to $Q$, which allows us to relate $j(\omega_{\sss C})$ to $\omega_{\sss E}$,
can be obtained by setting the coordinate on the fiber $H$ of $P_{\sss U}$ to the neutral element $e$ and identifying the result with 
$Q_{\sss U}$:  
\begin{align*}
    U \times\{e\} \times G = Q_{\sss U}.
\end{align*}
This corresponds to the quotient operation performed in defining $Q$ as the associated bundle 
$P \times_{\sss H} G$ (see Appendix \ref{App-AssocBundleQ} for details). Performing this quotient in Eq.~(\ref{EQ-j-matrix-Triv}) gives 
$\left(\varphi^*\right)^{-1}(\omega_{\sss C})$, the Ehresmann connection we are looking for, in the local 
trivialization corresponding to $Q_{\sss U}$:

\begin{equation}\label{EQ-OmegaEhresTriv}
    \omega_{\sss E}(x,g) = g^{-1}\omega^{\sss U}_{\sss C}(x) g + g^{-1} dg.
\end{equation}

Now, observe that Eq.~(\ref{EQ-OmegaLocalTriv}) also applies, in particular, to the Ehresmann connection $\omega_{\sss E}$ in the 
trivialisation corresponding to $Q_{\sss U}$.Then, comparing  Eq.~(\ref{EQ-OmegaLocalTriv}) for $\omega_{\sss E}$ with the above 
Eq.~(\ref{EQ-OmegaEhresTriv}) leads to:
\begin{equation}\label{EQ-OmegaEhresU=OmegaCartU}
   \omega^{\sss U}_{\sss E}(x) = \omega^{\sss U}_{\sss C}(x)
\end{equation}

\subsection{Levi-Civita coupling from Weitzenb\"ock one-form}\label{SUBSEC-jModifContorsion}
The above considerations show us that, when restricted to the base manifold, both the original Cartan connection and the Ehresmann connection 
obtained from it, thanks to 
Sharpe's theorem \ref{THM-Sharpe}, are the same on the base manifold, see 
Eq.~(\ref{EQ-OmegaEhresU=OmegaCartU}). They differ mainly through 
the Maurer-Cartan form between Eqs.~(\ref{EQ-OmegaEhresTriv}) and (\ref{EQ-OmegaCartanTriv}). In 
particular, for the
reductive Cartan connection of Eq.~(\ref{EQ-CartanReducConnec-W+thet}),
the Weitzenb\"ock term remains untouched by the map 
$(\varphi^*)^{-1}$, which thus cannot lead to a Levi-Civita coupling
.

To remedy this problem, one starts by observing that any Ehresmann connection is related to any 
other by a $G$-invariant\footnote{By $G$-invariance we mean the property for a one-form $\alpha$ to satisfy: 
$R_g^*\alpha = Ad_{g^{-1}} (\alpha)$, that is for matrix Lie group $R_g^*\alpha =  g^{-1} \alpha g $.} horizontal $\mathfrak{g}$-valued 
one-form. 
This can be seen as follows. First, let us consider the difference of any pair of Ehresmann connection one-forms 
$\omega_1$ and $\omega_2$. Using Eq.~(\ref{EQ-OmegaLocalTriv}) repeatedly for $\omega_1$ and $\omega_2$, and substracting the 
result one obtains, in the same local trivialization
\begin{equation*}
    \omega_2 - \omega_1 = g^{-1} \left(\omega^{\sss U}_2 - \omega^{\sss U}_1\right) g.
\end{equation*}
This $\mathfrak{g}$-valued one-form is manifestly $G$-invariant and horizontal. 
Second, if $\alpha$ is a $G$-invariant horizontal  $\mathfrak{g}$-valued one-form in $Q$, the 
sum $\omega + \alpha$, where $\omega$ is an Ehresmann connection, is both  $G$-invariant and $\mathfrak{g}$-valued. In addition,
since $\alpha$ is horizontal, $\omega + \alpha$ reduces to the Maurer-Cartan form along fibers. 
Thus  the $\mathfrak{g}$-valued one-form $\omega + \alpha$ is $G$-invariant, and reduces to the Maurer-Cartan form along 
the fibers, it is consequetly an Ehresmann connection one-form.

This property allows us to recast Eq.~(\ref{EQ-OmegaEhresTriv}), up to a redefinition of $\omega_{\sss E}$ under the form
\begin{equation} \label{EQ-SharpeOmegaModMatrixTriv}
\left(\omega_{\sss E} + \alpha\right)(x, g) = g^{-1} \left(\omega^{\sss U}_{\sss C} \right)(x) g + g^{-1} dg,
\end{equation}
\noindent where $\alpha$ is a $G$-invariant horizontal  $\mathfrak{g}$-valued one-form in $Q$. The Eq.~(\ref{EQ-OmegaEhresU=OmegaCartU}) in
the trivialization corresponding to $Q_{\sss U}$  becomes accordingly
\begin{equation*}
   \omega^{\sss U}_{\sss E}(x) + \alpha^{\sss U}(x) = \omega^{\sss U}_{\sss C}(x).
\end{equation*}
Then Eq~(\ref{EQ-SharpeOmegaModMatrixTriv}) can be recast under the form
\begin{equation} \label{EQ-SharpeOmegaModMatrixTriv-2}
\omega_{\sss E}(x, g) = g^{-1} \left(\omega^{\sss U}_{\sss C} - \alpha^{\sss U} \right)(x) g + g^{-1} dg,
\end{equation}

Now, let us specialize to our framework in which $P\nolinebreak=\nolinebreak OM$, $Q=AM$ and $\omega_{\sss C} = \omega_{\sss W} + \theta$. In that 
case, one can show \cite[theorem 6.2.5 p. 79]{Bleecker:1981} that, for a given Ehresmann one-form $\omega$ on $P = OM$, 
there exist a unique one-form $\kappa_\omega$  on $P = OM$, the so-called contorsion one-form, such that $\omega -  \kappa_\omega = \omega_{\sss LC}$, 
the Levi-Civita one form. The contorsion $\kappa_\omega$ being thus the difference between two Ehresmann connections 
it has the properties required to enter in Eq.~\nolinebreak(\ref{EQ-SharpeOmegaModMatrixTriv-2}) as the one-form $\alpha$. 
We can therefore set $\alpha$ to the contorsion corresponding to the Weisenb\"ock connection 
\begin{equation*}
    \alpha = \kappa_{\omega_{\sss W}} =:\kappa_{\sss W}
\end{equation*}
in order to obtain the Levi-Civita one-form in the  Ehresmann connection $\omega_{\sss E}$ when the Ehresmann part of the 
reductive Cartan connection $\omega_{\sss C}$ is $\omega_{\sss W}$.
Finally Eq.~(\ref{EQ-SharpeOmegaModMatrixTriv-2}) specializes to

\begin{equation} \label{EQ-SharpeOmegaModMatrixTriv-3}
\omega_{\sss E}(x, g) = g^{-1} \left((\omega_{\sss W} 
+ \theta)^{\sss U} - \kappa_{\sss W}^{\sss U} \right)(x) g + g^{-1} dg,
\end{equation}


\subsection{The $AM \longrightarrow OM$ map, and the covariant derivative}\label{SUBSEC-MapAM-OM-DefCovDer}
In the two previous sections we have shown how, starting from the Cartan-Weitzenb\"ock one-form (\ref{EQ-CartanReducConnec-W+thet}) 
in $P\nolinebreak=\nolinebreak OM$, one can obtain the Ehresmann-Levi-Civita one-form $\omega_{\sss E} = \omega_{\sss LC} + \theta$  in $Q=AM$, 
the principal Poincar\'e bundle. 

As a last step, the covariant derivative, corresponding to the Levi-Civita connection, over the base manifold $M$, can now be 
obtained thanks 
to a theorem shown in \cite[proposition 3.1 p. 127]{KobayashiNomizu:1963}  which states\footnote{We specialize here this 
theorem to 
the sub-bundle of orthonormal frame $OM$ and its affine extension $AM$.} the existence of a map, hereafter $\beta$, which associates to 
an affine connection, 
generically  $\omega + \phi$ defined on $AM$, the pair $(\omega, \phi)$ on $OM$. 

This applies in particular to the affine connection $\omega_{\sss E}$ 
given in Eq.~(\ref{EQ-SharpeOmegaModMatrixTriv-3}) for which   
$\beta: \omega_{\sss E} \mapsto (\omega_{\sss LC}, \theta)$. 
This map allows us to define the covariant derivative, associated to 
the reductive Cartan connection (\ref{EQ-CartanReducConnec-W+thet}), as 
the usual covariant derivative of GR, that is the Fock-Ivanenko derivative.
This is the main result of this section.

However, let us emphasize that the Levi-Civita connection appearing in this covariant derivative 
should be considered as a function of $\omega_{\sss W}$, 
$\theta$ and $\eta$ (the Minkowskian metric) these two last quantities entering in  the definition 
of the contorsion $\kappa_{\sss W}$. As a consequence, the Levi-Civita connection one-form 
should not be associated to the gauge field mediating gravity.

\section{Viewing TEGR as a gauge theory of translations ?}\label{SEC-TEGRAsTrans}

\subsection{Gauge field vs. connection}\label{SUBSEC-gf-vs-connec}
In gauge theories of particle physics, the gauge fields (associated with gauge bosons) are those fields which, at our present energy scale, 
mediate one of the 
fundamentals interactions (electromagnetic, weak or strong) between matter fields. They are termed gauge field since their dynamical free equations 
(uncoupled from matter), involving gauge fields through their field strength, exhibit gauge invariance. 
On the mathematical side, the gauge fields are recognized to be sections of Ehresmann connections
defined on a principal bundle, whose structure group $G$ is a global symmetry group of the free (in the sense of uncoupled through gauge fields) matter equations. 
The field strengths are (sections of) the curvature of these connections one-form.
The coupling between a matter field and a gauge field, renders the interacting theory of matter field locally invariant under 
the symmetry group $G$. Thus, in these theories, on the physical side, the gauge field is a dynamical field which fulfills two related roles: 
to mediate an interaction and to ensure local invariance under some symmetry. 

In classical gravity the spacetime is a metric manifold $(M, g)$, the metric being, in the Cartan view, induced by orthonormal (co-)frames (tetrads)
 through ${\displaystyle 
 \eta(e,e) = g}$. This manifold is canonically the base of a frame bundle $FM$. 
It is a Gl$(4,\setR)$ 
principal bundle which contains the orthonormal bundle $OM$ as a principal SO$(1,3)$ sub-bundle. A specific structure, 
the canonical one-form $\theta$,  
is always defined on $FM$. It realizes the so-called soldering\footnote{Note that $\theta$ is not the solder form by itself \cite[see][]{Fontanini:2018krt}.} 
and is independent of the existence of any connection on $FM$. In particular, it is worth noting that $\theta$
is not a connection one-form by itself. When an Ehresmann connection is present on $FM$, the canonical one-form allows us to define the torsion. 
The one-form $\theta$ is specific of the frame bundle, with no equivalent in the mathematical framework of particle physics gauge theory 
just described, where the Frame bundle, although always defined, is not used. As a consequence, the particle physics framework can be expected to be too restrictive to encompass a gauge theory of gravity involving torsion, such as TEGR.

The above remarks lead us to consider the role played by $\theta$ in our proposal to describe TEGR with the help of the reductive Cartan connection 
$\omega_{\sss C} = \omega_{\sss W} + \theta$. The canonical one-form appears in two places: 
\begin{enumerate}
\item in the definition of $\omega_{\sss C}$ Eq. (\ref{EQ-CartanReducConnec-W+thet}), where 
as a one-form valued in $\setR^4$, the translation part of the Poincaré algebra, it is identified with 
the  term $\theta$.  
\item in the definition of the contorsion one-form used to rewrite 
the Levi-Civita one-form as the combination 
$\omega_{\sss LC} = \omega_{\sss W} - \kappa_{\sss W}$  (see Sec.~\ref{SUBSEC-MapAM-OM-DefCovDer}).
\end{enumerate}

Thus, we observe that  the canonical one form $\theta$ is, first, the part of the Cartan connection related to the local translation invariance, 
and second, enters in an essential way in the coupling with matter. These two characteristics are reminiscent of those retained at the beginning of the present section to identify a gauge field. Since such a field is defined on the base manifold, let us examine the pullback, on the 
base manifold, of the canonical one-form and its associated Levi-Civita connection, 
that is the connection that enter 
the usual gravitational covariant derivative on the base. In the present context, 
the pullback along some section $\sigma$ of the canonical form $\theta$ reads 
\begin{equation}\label{EQ-link-theta-e}
\sigma^*\theta = e,
\end{equation}
where $e$ is a local field of frame (a tetrad), and that of
the connection $\omega_{\sss LC}$, in some open set $U$ of the base, reads 
\begin{equation}\label{EQ-Fake-LC-base}
\omega_{\sss LC}^{\sss U} = \omega_{\sss W}^{\sss U} - \kappa_{\sss W}^{\sss U}.
\end{equation}

The Eq.~(\ref{EQ-link-theta-e}), will play a central role in the translation-gauge interpretation,  because it 
relates $\theta$, a canonical structure, 
to the dynamical field $e$. Indeed, Eq.~(\ref{EQ-link-theta-e}) can be read of as the one-to-one relation between $\sigma$ and $e$ 
induced by $\theta$: to choose a section is to choose a frame. Now, $e$, in the Cartan view of gravity, is the solution of the 
gravity field equations. These, are known to exhibit gauge invariance, the choice of a gauge, that is the choice of a particular frame $e$, 
being, in the fiber bundle context, precisely the choice of a section $\sigma$. These facts point towards the interpretation of the 
frame $e$ as the gauge field of the theory.

Eq.~(\ref{EQ-Fake-LC-base}), on another hand can be viewed as the definition of a ``fake gauge field'', in the sense 
that the Levi-Civita term corresponds to the implementation of the local Lorentz invariance through the covariant derivative, 
but has no proper dynamics, other than that being induced by the tetrad $e$. 
Indeed, the first term on the r.h.s. of (\ref{EQ-Fake-LC-base}), the Weitzenb\"ock connection one-form, only involves local Lorentz
transformations and has a null curvature (field strength), while the second term on the  r.h.s. of (\ref{EQ-Fake-LC-base}), the contortion one-form,
is built on $\eta$, the constant Minkowskian metric, and the field $e$. The tetrad $e$ thus appears here as a dynamical field which ``drives'' the Lorentz invariance.

We finally come to the conclusion that, at least in the context of TEGR, a distinction should be made between the gauge field and the connection.
Note that, such a distinction  does not imply any change in particle physics theory (at least at our energy scale), since the structures 
involved in that extension are not present (not used) in the particle physics framework.


\subsection{A new gauge paradigm for TEGR ?}
This observation, shows us a possible way to interpret TEGR as a gauge theory of translations if we allow one to broaden the structure 
of a gauge theory by introducing a distinction between the gauge field and the connection. In this view, the gauge field is defined as the pullback 
on the base manifold, along some section $\sigma$, of the canonical form $\theta$, that is as a frame $e$. To remind the motivations
for such an interpretation, let us recall that the field $e$:
\begin{enumerate}
\item is a dynamical field, whose equation exhibits gauge invariance,
\item mediates the interaction
through the Levi-Civita connection, Eq.~(\ref{EQ-Fake-LC-base}), which in the present context
is induced by the field $e$, as described at the end of 
Sec.~\ref{SUBSEC-gf-vs-connec},
\item 
is to some extent responsible for the local Lorentz invariance in the sense that the Levi-Civita
connection is here determined by $e$ and structural elements as $\eta$. 
\end{enumerate}
These three properties are characteristic of a gauge field (Sec.~\ref{SUBSEC-gf-vs-connec}). Then, if we insist to interpret 
TEGR as a gauge 
theory, we can describe it using $e$ as the gauge field of translations associated with the 
Cartan connection $\omega_{\sss C}$ Eq.~(\ref{EQ-CartanReducConnec-W+thet}). 

As explained at the end of Sec.~\ref{SUBSEC-gf-vs-connec}, this interpretation does not require any changes in the usual framework of particle
physics gauge theories since the departure from that framework relates to extraneous quantities: $\theta$, $e$. Nevertheless, it requires 
a distortion of the attributes of the gauge field, in the sense that the field associated
to translations, the tetrad $e$, does not implement a local invariance nor mediates the interaction in the same way as gauge field of particle physics would do. This is of course related to the nature of that field, which is not a connection. 

Here, the coupling made through the Levi-Civita connection, 
although induced by the translation field $e$, relates to Lorentz invariance. Furthermore, coupling to matter involves the corresponding representation 
of the Lorentz group, in particular matter's spin. For a scalar field, as it is spinless, that coupling reduces to zero. 
Since gravity seems to couple universally to matter, that is independently of its spin, 
the gravitational coupling should also arise from elsewhere. Indeed, since the generators of the translations span the space 
$\mathfrak{g}/\mathfrak{h} = \setR^4$ of the Poincar\'e algebra and, as seen in Sec.~\ref{SUBSEC-CartanConnec}, this space is identified 
to the tangent space of the base manifold, the expansion of the partial derivative operator on the tetrad basis in a neighborhood of some point $x$,   
\begin{equation*}
    \partial_\mu = \left(\partial_\mu\right)^a\, e_a (x) 
\end{equation*}
shows that the partial derivatives are related to local translations. The universal coupling to gravity should thus be assigned to 
the partial differential operator, and related to local and infinitesimal translations. We remark that this interpretation is reminiscent
of that of the translation-only gauge theory \cite[see][sec 5.3]{Aldrovandi:2013wha}, although it avoids the problematic identification of the gauge field with a connection
mentioned in the introduction Sec.~\ref{SEC-Intro}.

Here, we point out that, although the structure of the coupling to matter of our 
Cartan-TEGR formulation is mathematically well defined, the gauge interpretation of the tetrad is more a matter of opinion
. The conceptual split between the concepts of gauge field and connection being sound, objections can be raised on the protracted reasoning that leads to the link between translations and the gauge field $e$.
In particular, gauging translations in the present context does not correspond to replacing a global symmetry by a local one, the translations generated by the $\mathfrak{g}/\mathfrak{h}$ part of the Poincar\'e algebra being always local (and infinitesimal). 
 We offer such interpretation to the adhesion 
 from the reader, but abstain from claiming it.


\section{Conclusion}\label{SEC-Conclu}


The main aim of this article is to show that it is possible to retrieve the correct coupling to matter in TEGR starting from a Cartan connection and following a well defined and robust procedure to obtain the familiar Levi-Civita form which of course fits all presently available data. To achieve this we use a powerful theorem by Sharpe that yields a one to one correspondence between Cartan connections and Ehresmann affine connections, we apply the Equivalence Principle to extract the Levi-Civita connection from TEGR's Weitzenb\"ock one-form, we eventually map the results from the Affine bundle to the Frame bundle. Following these steps the coupling to matter is then given by the usual Fock-Ivanenko covariant derivative appearing as a consequence of the structure descending from the initial choice of Cartan connection. 
 

Beside showing how to coherently retrieve the correct coupling to matter from TEGR with a Cartan connection, we adventure in discussing how the usual paradigm of gauge theories in classical particle physics
needs to be enlarged if one insists on interpreting the structure of the TEGR as a gauge theory for the translation group. More precisely, it appears that a distinction
between the connection (on spacetime) and the gauge field is required: in such interpretation torsion appears as the field strength 
of the Cartan connection, justifying the Cartan structure to reproduce TEGR in a bundle framework, and the co-frame (tetrad) as the gauge field related to local translation invariance. We note furthermore that the Lorentz invariance is also driven, although indirectly,  by the co-frame.  

As an interpretation of the theory here described, the gauge theoretic nature of TEGR could certainly still be discussed and adhesion to such interpretation is left to the reader. Nevertheless, 
the Cartan connection approach gives a new and consistent theoretical description of TEGR.

\section*{Acknowledgements}

The authors wish to thanks, D. Bennequin for helpful discussions on geometry. The work of M.~Le~D. has been supported by 
Lanzhou University starting fund, and the Fundamental Research Funds for the
Central Universities (Grant No.lzujbky-2019-25).

\medskip
\appendix
\section{Relating Ehresmann and Cartan connections}\label{App-EhressToCart}
For convenience we reproduce here the theorem  \cite[Prop. 3.1 p. 365 of][]{Sharpe:1997}, relating Ehresmann to Cartan connections, that we use to obtain a 
covariant derivative. 

Let $G$ be a Lie group and $H$ a subgroup of $G$, \linebreak
$P:=\sbundle{P}{M}{H}$ a principal bundle. 

Let $Q:=\sbundle{Q}{M}{G} = P \times_{\sss H} G$ 
the principal $G$-bundle associated to $P$ by the action by left 
multiplication of $H$ on $G$. The principal 
bundle $P$ is a sub-bundle of the principal bundle $Q$
through the canonical inclusion 
$p\nolinebreak\mapsto\nolinebreak(p,e_{\sss G})$.

Let $E_{\sss Q}$ be the set of Ehresmann 
connection $\omega_{\sss E}$ on $Q$ such that Ker$({\omega_{\sss E}})\bigcap\varphi_*(TP)= 0$, 
and $C_{\sss P}$ the set of $\mathfrak{g}$-valued Cartan connections on $P$.

\begin{thm}\label{THM-Sharpe}
 Let $(G,H)$ be a Klein geometry\footnote{A Klein geometry is a pair $(G,H)$, where $G$ is a Lie group and $H \subset G$ a 
closed subgroup such that $G/H$ is connected \cite[see][p. 151]{Sharpe:1997}} and 
let $P$ and $Q$ be principal $H$ and $G$ bundles,  
over a manifold $M$, respectively. 
Assume that dim $G = $ dim $P$ and that $\varphi: P \mapsto Q$ 
is an $H$-bundle map. Then the correspondence  
$\varphi^*: E_{\sss Q} \longrightarrow C_{\sss P}$, 
is a bijection of sets.
\end{thm}

The inverse map of $\varphi^*$ 
is defined as follow: 
let $\omega_{\sss C}$ a $\mathfrak{g}$-valued Cartan connection on $\sbundle{P}{M}{H}$,
it can be extended to 
a one-form $j(\omega_{\sss C})$ on $P \times G$  defined by the expression: 
\begin{equation}\label{EQ-def-j}
   j(\omega_{\sss C}) := Ad_{g^{-1}} \pi_{\sss P}^* \omega_{\sss C} + \pi_{\sss G}^* \omega_{\sss G},
\end{equation}
where $\omega_{\sss G}$ is the Maurer-Cartan form on $G$, while $ \pi_{\sss P}\textrm{ and } \pi_{\sss G}$ are the projections on $P$ and $G$ respectively.
This one-form on $P \times G$, is proven to be the pull-up, from $Q$ to $P \times G$, of the Ehresmann connection $\omega_{\sss E}$ such that 
$\varphi^*(\omega_{\sss E}) = \omega_{\sss C}$.





\section{Associated bundle $Q$}\label{App-AssocBundleQ}

In Ref.~\cite{Fontanini:2018krt}, we commented on associated bundles, and in particular, on associated vector bundles. Here we recall some facts about 
associated Lie group bundles in relation to our application of the Sharpe theorem to the Cartan connection in Sec.~\ref{SUBSEC-SharpeTheorem}
. 

Let us remind from \cite[appendix 4]{Fontanini:2018krt} that\footnote{We use here the notation $\times_{\sss H}$ instead of $\times_\rho$
in coherence with that of Sharpe and \cite{Isham:1999qu}.} $P \times_{\sss H} F$, where $P=\pbundle{P}{M}{H}{\pi}$ is a principal 
left $H$-bundle and $F$ a $H$-space, is a manifold whose points are the orbits (the equivalence
classes) for the right action $R_{\sss H} (p, f) \mapsto R_{\sss H}(p, f) := (ph, h^{-1}f)$ of $H$ on the product space $P\times F$. 
The projection map $\pi$ of $P$ induces a projection $\hat \pi$ from $P\times_{\sss H} F$ onto the base $M$. The fiber of 
$P\times_{\sss H} F$ over some $x\in M$ is then defined as $\hat\pi^{-1}(x)$. Then, on can show that the local differentiable structure of 
$P$ ensures that $P\times_{\sss H}F$ is a fiber bundle with base $M$, fiber $F$, and structure group $H$. 

In Sharpe theorem, the resulting associated bundle $Q = P \times_{\sss H} G$ is, in fact, a principal $G$-bundle. This is because 
the Lie group $G$ (the Poincar\'e group in our particular framework) contains the Lie group $H$ (the Lorentz group) as a subgroup.
Indeed, $Q$ is the so-called $G$-extension of $P$ \cite[see][Sec. 5.3.3]{Isham:1999qu}. As a consequence, following our notations 
Sec. \ref{SUBSEC-SharpeTheorem}, a local trivialisation
of $Q = P \times_{\sss H} G$  correspond to the local product
$U\times G$, $U$ being some open set of trivialization, corresponding
local coordinates on $Q$ are $(x, g)$.


\bibliography{TEGRbiblio}
\end{document}